# Working Memory Capacity & Gender: Small Overall Differences between Genders, U-Shaped Curve for Male/Female Ratio

Regina Ershova[1] and Eugen Tarnow[2]

[1]Department of Psychology, State University of Humanities and Social Studies (SUHSS).
Zelenaya str., 30, Kolomna, Russia, 140410

[2]Avalon Business Systems, Inc.
18-11 Radburn Road, Fair Lawn, NJ 07410, USA

etarnow@avabiz.com

## Abstract

The working memory capacity (WMC) of 400 Russian college students was measured using the Tarnow Unchunkable Test [2] which tests WMC alone without requiring explicit working memory operations.  We found small-sized WMC differences by gender and a u-shaped curve: the male/female ratio increases for low and high WMC.  The gender proportion in each academic fields was a strong determinant of the average WMC ($r^2$=0.2 for the 3-item test and $r^2$=0.5 for the 4-item test), associating "academic female" (law, history) with holistic thinking and "academic male" with reductive thinking (physics, computer science, math). Within academic fields there were no WMC gender differences. The male/female ratios for the different fields are strongly amplified from the WMC male/female ratios, by factors of 12-14.

NEED GENDER RATIOS OF THE STUDENT ADMISSIONS

**Introduction**

If the reader attempts to remember four unrelated double digit integers (the Tarnow Unchunkable Test, TUT [2]), it is very probable that one of the integers vanishes, no matter how hard the reader tries. This almost magical process shows that most of us have three memory slots for similar items. If we try to force in another item into our "working memory", interference occurs and removes one of the other items - somehow.

We distinguish between two types of working memory capacity: working memory capacity measured in tasks that also include a variety of operations such as arithmetic problem solving or sentence content analysis which we call working memory operational capacity (WMOC) as opposed to working memory capacity measured without such operations which we call working memory capacity (WMC).

WMOC is thought to be strongly correlated with general fluid intelligence and novel reasoning, perhaps via attentional control (Kane et al, 2005). It is found that WMOC can be lowered by math anxiety (Ashcraft & Kirk, 2001; Miller & Bixel, 2004) and stereotype threat (Schmader & Johns, 2003) and sometimes increased by physical exercise (Sibley & Beilock, 2007).

WMC is measured differently by different researchers, some use simple free recall as a measure. Free recall consists of two stages, only the first corresponds to working memory, the second is a reactivation process presumably using associations (Tarnow, 2015). In this contribution we use the TUT which attempts to eliminate the possibility of associations and the activation of the second stage.

In general, gender differences vary from very large (height), large (some preferences) to small (cognitive properties), see Hyde (2014). In this contribution we investigate how WMC varies with gender which apparently has not been done before CHECK.

The results of our investigation may be important for understanding gender differences, the structure of WM, for designing new empirical studies to advance theory and research in this area, as well as measuring the effectiveness of the methodological tools needed to test WM.

## Method

We present data from a study of university students aged 17 to 24.

The Tarnow Unchunkable Test (TUT) used in this study separates out the working memory (WM) component of free recall by using particular double-digit combinations which lack intra-item relationships (Tarnow, 2013). It does not contain any explicit WM operations. The TUT was given via the internet using client-based JAVAScript to eliminate any network delays. The instructions and the memory items were displayed in the middle of the screen. Items were displayed for two seconds without pause. The trials consisted of 3 or 4 items after which the subject was asked to enter each number remembered separately, press the keyboard enter button between each entry and repeat until all the numbers remembered had been entered. Pressing the enter button without any number was considered a "no entry". The next trial started immediately after the last entry or after a "no entry". There was no time limit for number entry. Each subject was given six three item trials and three four item trials in which the items are particular double-digit integers.

## Sample

193 Russian undergraduate students of the State University of Humanities and Social Studies (121 (63%) females and 71 (37%) males, mean age was 18.8 years) participated in the study for extra credit. Each participant was tested individually in a quiet room. An experimenter was present throughout each session.

One record was discarded – the student had only responded once out of a possible thirty times.

**Results**

There is no WMC gender difference in the average or variance for the 3-item test (one-way ANOVA yields F=0.6, p=0.44; F-test p=0.4) but there is a small difference for the 4-item test (one-way ANOVA yields F=7.5, p=0.006, Cohen's d=0.26; F-test p=0.046; where we have adopted Hyde's (2014) definition:

d=(mean score of males - mean score of females)/(mean(within-groups standard deviation)).

The 4-item difference is similar in size to other cognitive effects for gender, see Hyde, 1981 and Hyde, 2014), see Fig. 2. If we plot the ratio of males to females with a particular WMC (Fig 3), we find u-shaped curves; there are more males at high and low WMC than there are females for both tests. The variance ratios (Hyde, 2014) we find to be 1.05 for the 3-item test and 1.26 for the 4 item test.

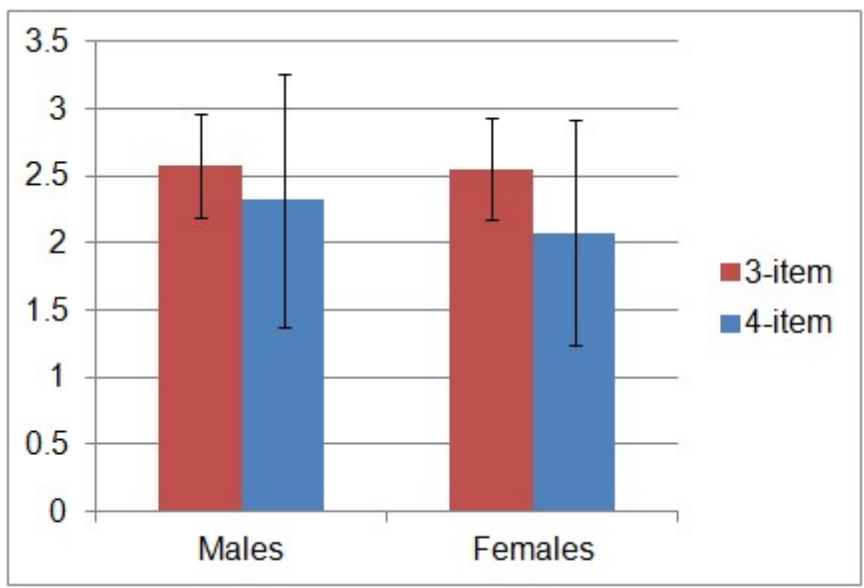

*Figure 1. Average recall as a function of gender.*

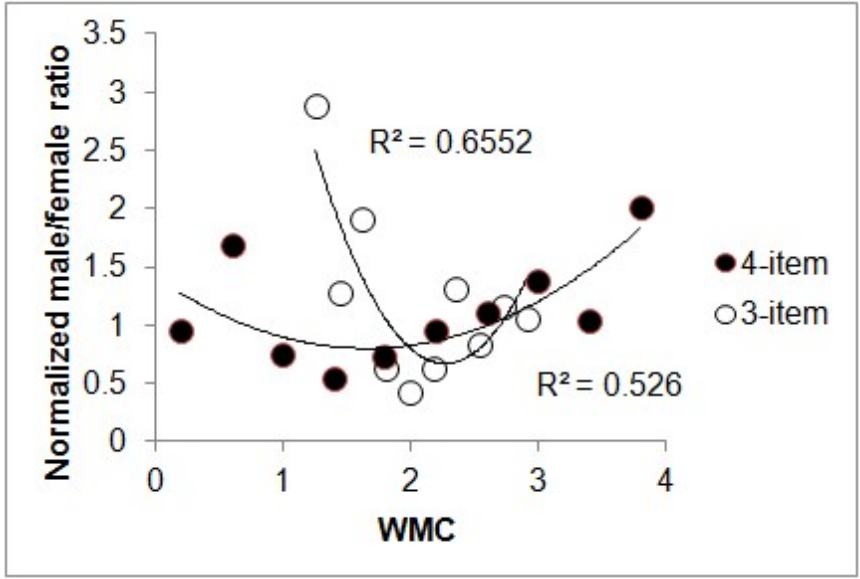

*Figure 2. Ratio males to females as a function of WMC for 3-item and 4-item recall indicate u-shaped curves.*

Using Bayesian classification, random predictions yielded 69% for gender. Using 3-item WMC and 4-item WMC we get 71%.

WMC scores depend heavily on academic field (Ershova & Tarnow, 2017), dwarfing the overall gender differences in WMC and the WMC standard deviation, see Fig. 3.

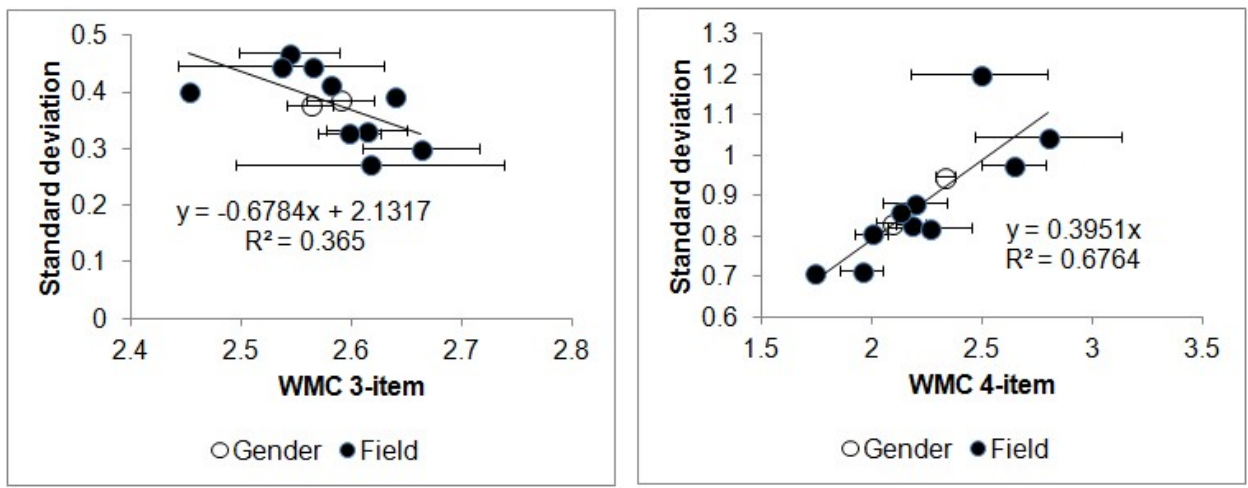

*Figure 3. WMC and the corresponding deviations as a function of average recall of the different fields (filled circles) and gender (unfilled circles).Gender differences mimic differences between different academic fields but are a lot smaller.*

Figure 4 displays the memory scores for each field as a function of the gender balance of that field, which we term "academic gender". We find that fields with more male students had students with higher WMC than fields with more female students and these are large sized effects. If we remove the field dependency there is no gender differences within the fields (F=1.7 p=0.19 for 3-item test; F=0.5 p=0.41 for 4-item test). The WMC by gender and field is displayed in Fig. 4. If we replot the results of academic gender as in Fig. 2, we obtain the results of Fig. 5. For the 3-item test there is no u-shaped curve because of its limited range. For the 4-item test the differences are amplified in Figure 5 compared to Figure 2, perhaps suggesting that the cause and effect may be reversed.

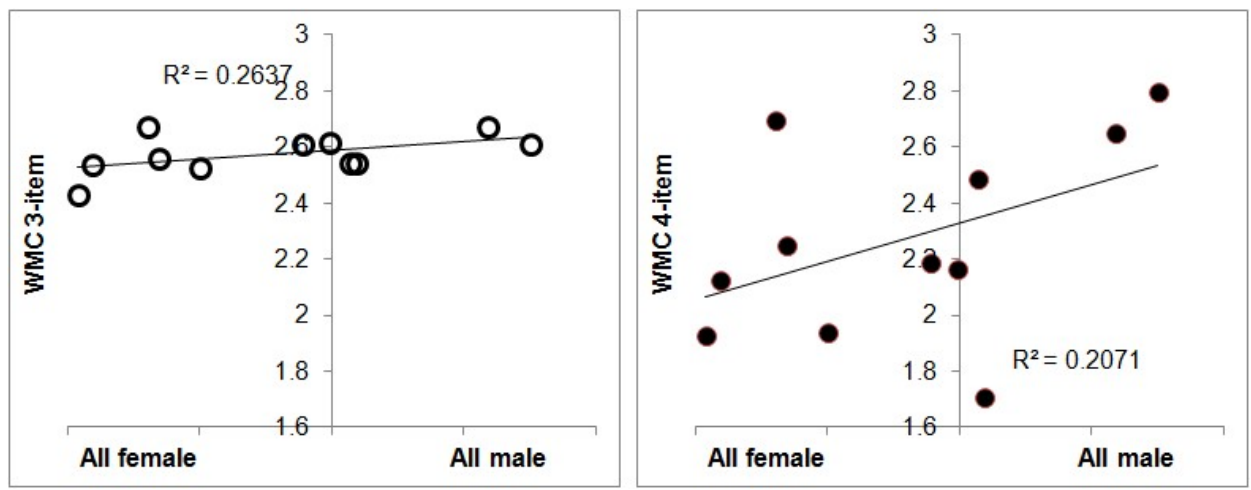

*Figure 4. Average recall of a field as a function of the average gender of the field. Left is the 3-item test, right is the 4 item test.*

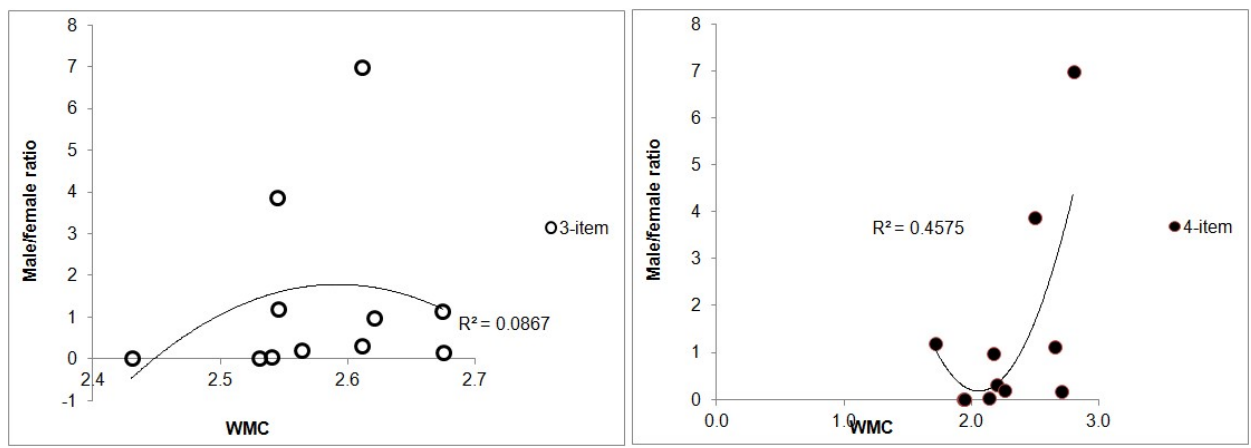

*Figure 5. Field male/female ratio versus WMC. Comparison with Fig. 2 suggests that for the 4-item test, the male/female ratios are greatly amplified. For the 3-item test there is no correlation.*

**Discussion**

We found small differences in WMC between the genders and they are dwarfed by the effects of the academic field. Males are more common than females among low and high WMC.